\documentclass[preprint,12pt]{elsarticle}
\usepackage[centertags]{amsmath}
\usepackage{amsfonts}
\usepackage{amssymb}
\usepackage{amsthm}
\usepackage{pgf}
\usepackage[T1]{fontenc}
\usepackage{indentfirst}
\usepackage{mathrsfs}
\usepackage[ansinew]{inputenc}
\usepackage{multirow}

\journal{Physica A: Statistical Mechanichs and its Applications}

\begin{document}




\title{A new parameterization for the concentration flux using the fractional calculus to model the dispersion of contaminants in the Planetary Boundary Layer}


\author[]{A. G. Goulart \corref{mycorrespondingauthor}}
\ead{antonio.goulart@furg.br}
\cortext[mycorrespondingauthor]{Corresponding author}

\author[]{M.~J. Lazo }

\author[]{J. M. S. Suarez }

\address[]{Instituto de Matem\'atica, Estat\'{\i}stica e F\'{\i}sica, Universidade Federal do Rio Grande, 96.201-900 Rio Grande, Rio Grande do Sul, Brazil.}

\begin{abstract}
In the present work, we propose a new parameterization for the concentration flux using fractional derivatives. The fractional order differential equation in the longitudinal and vertical directions is used to obtain the concentration distribution of contaminants in the Planetary Boundary Layer. We solve this model and we compare the solution against both real experiments and traditional integer order derivative models. We show that our fractional model gives very good results in fitting the experimental data, and perform far better than the traditional Gaussian model. In fact, the fractional model, with constant wind speed and a constant eddy diffusivity, performs even better than some models found in the literature where it is considered that the wind speed and eddy diffusivity are functions of the position. The results obtained show that the structure of the fractional order differential equation is more appropriate to calculate the distribution of dispersed contaminants in a turbulent flow than an integer-order differential equation. Furthermore, a very important result we found it is that there should be a relation between the order $\alpha$ of the fractional derivative with the physical structure of the turbulent flow.

\end{abstract}



\maketitle

\section{Introduction}

Atmospheric turbulence plays a fundamental role in the dispersion of contaminants in the Planetary Boundary Layer (PBL). In fact, without turbulence, the contaminants would follow only the streamlines of mean wind velocities displaying minimal diffusion in other directions. From the theoretical point of view, turbulence is a permanent source of challenging problems due to its physical and mathematical complexity. As an example of this complexity, a remarkable consequence of turbulence is the emergence of the anomalous diffusion phenomenon. Unlike common diffusion, where the mean square displacement increases linearly with time, in the anomalous diffusion, the mean square displacement is not linear. The anomalous diffusion is closely connected with the failure of the central limit theorem due to sparse distribution or long-range correlations. Actually, it is related to the more general L\'{e}vy-Gnedenko theorem that generalizes the central limit theorem for situations where not all moments exist \cite{Metzler2000}. Historically, anomalous diffusion was first observed in nature in the dispersion of contaminants phenomenon in 1926, when Richardson measured the increase in the width of plumes of smoke generated from point sources located in a turbulent velocity field \cite{Metzler2000, Metzler2004, Richardson, West,West2}.
Richardson speculated that the speed of turbulent air, which has a non-differentiable structure, can be approximately described by the Weierstrass function. This was motivated partly by the observation that the width of smoke plumes grows with $t^{\alpha}$ ($\alpha\geq 3$), unlike common diffusion where $\alpha=1$.

The non-differentiable behavior for the width growth of plumes generated from a point source is directly related to the fractal structure of the turbulent velocity field, where the fluctuation's size scales are, in many cases, very large compared to the average scale \cite{Mandelbrot}. In this context, the classical advection-diffusion equation does not fully explain the diffusion of contaminants since the parameters of the system usually grow faster than the solutions obtained by classical models \cite{West,West2}. Moreover, it is expected that traditional differential equations do not adequately describe the problem of turbulent diffusion because usual derivatives are not well defined in the non-differentiable behavior introduced by turbulence. Notwithstanding this fact, traditional Eulerian and Lagrangian models are the most frequently used in modeling the dispersion of contaminants in the atmosphere \cite{Gryning1987, Hanna and Paine, Moreira2005a, Wilson}. The Eulerian models consist basically of the advection-diffusion equation and the Lagrangian models are based on the  Langevin equation. In order to deal with the anomalous diffusion, a common procedure used in the literature in order to modify Eulerian models is to assume that the physical structure of turbulent flow and velocity fields are described by a complex eddy diffusivity and mean velocity profile that are both considered as functions of spatial coordinates. These functions or are usually chosen in order to fit experimental data or they are obtained from the Taylor statistical diffusion theory \cite{Taylor, Batchelor, Degrazia, Goulart2004}.

Despite the difficulties aroused by the non-differentiable behavior introduced by turbulence, in the last decades, few works dealt with the validity of models based on classical differential equations to describe systems displaying non-differential behavior and/or anomalous dynamics. In this subject, we can highlight the use of fractional calculus that emerged as a valuable mathematical tool to model the time evolution of anomalous diffusion (a detailed physical explanation for the emergence of fractional derivatives in anomalous diffusion models can be found in the reviews \cite{Metzler2000, Metzler2004,West,West2}. However, the use of fractional derivatives to study steady-state systems, and in particular the dispersion of contaminants in the atmosphere, is largely underexplored. In this context, we demonstrate in \cite{Goulart2017} that the steady-state regime for a spatial concentration distribution of a non-reactive contaminant, in an anomalous diffusion process displaying a power law mean square displacement, is given naturally by a differential equation containing fractional derivatives. Moreover, we show that the fractional model, even in the simplest case of a constant eddy diffusivity, gives very good results when compared with real data and models usually found in the literature with complex eddy diffusivity and mean velocity profile.

The promising result found in our previous work \cite{Goulart2017} motivates a deeper analysis of the relationship between the turbulence structure of the turbulent flow in the atmosphere and the anomalous diffusion expressed mathematically by a fractional diffusion equation. In the present work, our main objective is to make this deeper analysis by comparing the solution obtained from our fractional diffusion model with the Copenhagen experiment \cite{Gryning1984}, the Prairie Grass experiment \cite{Barad} and the Hanford experiment \cite{Nickola}. Furthermore, we also generalize the fractional differential model we introduced in \cite{Goulart2017} by considering a new parameterization for the concentration flux using the fractional calculus. In the work \cite{Goulart2017}, the parameterization for the concentration flux is the usual one, where it is assumed that the turbulence causes a net movement of material in the direction of the concentration gradient at a rate which is proportional to the magnitude of the gradient \cite{Pasquill}. However, the emergence of a fractional operator in the advective term due to the anomalous diffusion process \cite{Goulart2017} leads to a model with inconsistent space dimensions in its different terms. There are two simple ways to correct this inconsistency. The first is by the introduction of an additional scale parameter \cite{Metzler2000,Metzler2004,West,West2, Magin,Gomez-Aguilar}, and the second one is by the introduction of a fractional derivative in the diffusive term by an appropriate parameterization for the concentration flux. In the present work, we adopted the second approach by considering a parameterization for the concentration flux given by a fractional derivative (gradient) of the concentration. The physical justification for this approach is that the turbulent flow causes a net movement of material that reflects the fractal structure of the turbulent vortices, which can be related to the fractional order of the derivative present in the gradient of the flux \cite{West2,Tarasov}, motivating the use of this fractional parametrization to investigate the relationship between the turbulence structure of the turbulent flow in the atmosphere and the anomalous diffusion. We show that our fractional model with constant velocity field and a constant eddy diffusivity gives very good results when fitting the experimental data, and also when compared to more complex models with integer order derivatives. Most important, a result we found is that there should be a relation between the order $\alpha$ of the fractional derivative and the physical structure of the turbulent flow. Since, regardless of the experiment, when we have the predominance of mechanical energy input in the turbulent flow all experiments were best described with $\alpha = 0.72$, and when we have the predominance of energy input by thermal convection the experimental data is better described by $\alpha = 0.80$.

The paper is organized in the following way. In Section II we review the basic definitions and properties of Riemann-Liouville and Caputo fractional derivatives, that are needed for formulating our fractional differential equation model. The fractional model is presented and analytically solved in Section III. A numerical comparison of our solution against traditional models and experimental data is done in Section IV. Finally, Section V presents the conclusions.

\section{Model description}

An equation for the spatial distribution of concentration of a non-reactive contaminant in the PBL can be obtained by an application of the principle of continuity or conservation of mass \cite{Pasquill,Csanady,Seinfeld}:
\begin{equation}\label{eq 1}
 \frac{D \bar{c}}{D t}+\overrightarrow{\nabla}\cdot \overrightarrow{\Pi}_{c}=0,
\end{equation}
where $\frac{D}{Dt}=\frac{\partial}{\partial t}+\vec{u}\cdot \overrightarrow{\nabla}$ is the Lagrangian derivative, $\bar{c}=\bar{c}(x,y,z,t)$ is the average concentration, $\vec{u}$ is the wind velocity, and $\overrightarrow{\Pi}_{c}$ is the concentration flux.  A traditional closure for concentration flux problem \eqref{eq 1} is based on the gradient-transfer approach, which assumes that turbulence causes a net movement of material in the direction of the concentration gradient, at a rate which is proportional to the magnitude of the gradient \cite{Pasquill}:
\begin{equation}\label{eq 2}
 \overrightarrow{\Pi}_{c}=-K \overrightarrow{\nabla}\bar{c} ,
\end{equation}
where $K$ is the eddy diffusivity.
By replacing \eqref{eq 2} in \eqref{eq 1}, considering the steady-state case and choosing a Cartesian coordinate system in which the longitudinal direction $x$ coincides with the average wind velocity, and by neglecting the longitudinal diffusion, we get
\begin{equation}\label{eq 3}
 u\frac{{\partial}\;\overline{c}}{\partial x}-\frac{\partial}{\partial y}(K_{y}\frac{\partial \; \overline{c}}{\partial y})-\frac{\partial}{\partial z} (K_{z}\frac{\partial \; \overline{c}}{\partial z})=0.
\end{equation}
Finally, in order to compare the model with experimental data found in the literature, the equation for the cross-wind integrated concentration ($\overline{c^{y}}=\overline{c^{y}}(x,z)$) is obtained by integrating the Eq. \eqref{eq 3} with respect to $y$ from $-\infty$ to $+\infty$:
\begin{equation}\label{eq 4}
 u\frac{{\partial}\;\overline{c^{y}}}{\partial x}-\frac{\partial}{\partial z}(K_{z}\frac{\partial \; \overline{c^{y}}}{\partial z})=0.
\end{equation}

The Eq. \eqref{eq 4}  represents a classical model used to estimate the concentration distribution of a contaminant in the Earth's atmosphere, obtained by considering closures for concentration flux based on the gradient-transfer approach \eqref{eq 2}.

In the work \cite{Goulart2017}, the parameterization for the concentration flux is the usual one, given by \eqref{eq 2}, where it is assumed that the turbulence causes a net movement of material in the direction of the concentration gradient at a rate which is proportional to the magnitude of the gradient \cite{Pasquill}. However, the emergence of a fractional operator in the advective term due to the anomalous diffusion process \cite{Goulart2017} leads to a model with inconsistent space dimensions in its different terms. There are two simple ways to correct this inconsistency. The first is by the introduction of an additional scale parameter \cite{Metzler2000,Metzler2004,West,West2, Magin,Gomez-Aguilar}, and the second one is by the introduction of a fractional derivative in the diffusive term by an appropriate parameterization for the concentration flux. In this work, we adopted the second approach by considering a parameterization for the concentration flux given by a fractional derivative of the concentration,
\begin{equation}\label{eq 5}
 \Pi_{c,z}=-K_{z}\frac{\partial^{\beta} \; \overline{c}}{\partial z^{\beta}},
\end{equation}
where \cite{Diethelm, Goulart2017},
\begin{equation}
\label{eq 5a}
\!\!\!\!\! \frac{\partial^{\beta}f(x,y,z)}{\partial z^\beta} =\frac{1}{\Gamma(1-\beta)}\int_{a}^{z} \frac{\partial_{u}f(x,y,u)}{(z-u)^{\beta}}du,
\end{equation}
is the left Caputo partial fractional derivatives of order $\beta$ ($0<\beta < 1$) with respect to $z$,  $a\in \mathbb{R}$ and $\partial_u$ is the ordinary partial derivative of integer order with respect to the variable $u$ .
 The physical justification for this approach is that the turbulent flow causes a net movement of material that reflects the fractal structure of the turbulent vortices, which can be related to the fractional order of the derivative present in the gradient of the flux \cite{West2,Tarasov}. There are many works in the literature that suggest fractal models in order to describe turbulence (for examples, see \cite{Hentschel,Procaccia, Frisch}). In this context, fractional derivatives are in general more adequated to describe the problem than classical derivatives.


 On the other hand, we show in \cite{Goulart2017} that if the process displays anomalous diffusion with a power law mean squared displacement given by $\langle z^{2}\rangle \propto x^{\alpha}$,  with $0 <\alpha < 1$, then the order of the derivative in the advective term ($x$ direction) would be $\alpha$. In this case, the Eq. \eqref{eq 4} becomes,
\begin{equation}\label{eq 6}
 u\frac{{\partial^{\alpha}}\;\overline{c^{y}}}{\partial x^{\alpha}}=\frac{\partial}{\partial z}(K_{z}\frac{\partial^{\beta} \; \overline{c^{y}}}{\partial z^{\beta}}).
\end{equation}
where the left Caputo partial fractional derivatives of order $\alpha$ ($0<\alpha < 1$) with respect to $x$ is defined as in equation \eqref{eq 5a}. Note that for $\alpha=\beta=1$ \eqref{eq 6} reduces to the classical integer order diffusion equation \eqref{eq 4}, since \eqref{eq 6} follows from a direct generalization of \eqref{eq 4} for anomalous process \cite{Goulart2017}.

The fractional diffusion equation \eqref{eq 6} generalizes the model we introduced in \cite{Goulart2017} by the inclusion of a Caputo fractional derivative in $z$, as a consequence of the new closure for concentration flux introduced in Eq. \eqref{eq 5}.   A further advantage of our present Eq. \eqref{eq 6} against the model in \cite{Goulart2017} is that for $\alpha=\beta$ it has the same dimension on both sides of \eqref{eq 6}. Considering that \eqref{eq 6} is obtained from the law of conservation of mass, it is physically more correct to have the same dimensions on both sides of equality.  In this work, we will consider the simplest case for \eqref{eq 6}, when $u$ and $K_{z}$ are constants. In this case, the Eq. \eqref{eq 6} becomes,


\begin{equation}\label{eq 7}
 \frac{{\partial^{\alpha}}\;\overline{c^{y}}}{\partial x^{\alpha}}=\kappa \frac{\partial}{\partial z}(\frac{\partial^{\alpha} \; \overline{c^{y}}}{\partial z^{\alpha}}).
\end{equation}
where $\kappa = \frac{K_{z}}{u}$, and we consider also only the particular case where $\alpha=\beta$. In the particular case where $\alpha=1$, Eq. \eqref{eq 7} reduces to the well know advection-diffusion equation (Gaussian model) that describes a Gaussian process.

In order to \eqref{eq 7} describes a possible real dispersion process in PBL, it should be imposed boundary conditions of zero flux on the ground ($z=z_{0}$) and top ($z = h$), and consider that the contaminant is released from an elevated point source with emission rate $Q$ at height $H_{s}$, i.e.,
\begin{equation}\label{eq 8}
K_{z}\frac{\partial^{\alpha} c^{y} }{\partial z^{\alpha}}=0,\;\;\;\;z=z_{0},\;\;z=h,
\end{equation}
\begin{equation}\label{eq 9}
u c^{y}(0,z)=Q\delta(z-H_{s}),\;\;x=0,
\end{equation}
where $z_{0}$ is the surface roughness length, and $\delta(\cdot)$ is the Dirac delta function.

The solution of the fractional differential equation \eqref{eq 7}, subjected to the boundary conditions \eqref{eq 8} and \eqref{eq 9}, can be analytically obtained. We have (see \ref{appendix})
\begin{equation}\label{eq 13}
\overline{c^{y}}(x,z)=\sum_{n=0}^{\infty}a_{n}E_{\alpha}(-\kappa \lambda_{n}^{2}x^{\alpha})E_{\alpha+1}(-\lambda_{n}^{2}z^{\alpha+1}),
\end{equation}
where $E_{\alpha}(.)$ is the Mittag-Leffler function defined by,
\begin{equation}\label{eq 14}
E_{\alpha}(x)=\sum_{n=0}^{\infty}\frac{x^{n}}{\Gamma(n \alpha + 1)},
\end{equation}
and the constants $a_{n}$ and $\lambda_{n}$ are obtained from the boundary conditions \eqref{eq 8} and \eqref{eq 9} (see  \ref{appendix}).

The solution for the Gaussian model (given by \eqref{eq 7} with $\alpha=1$) is obtained from equation \eqref{eq 13} taking $\alpha = 1$,
\begin{equation}\label{eq 15}
\overline{c^{y}}(x,z)=\frac{Q}{u h}\Big[1+2\sum_{n=1}^{\infty}\cos(\lambda_{n}H_{s})\cos(\lambda_{n}z)\exp (- \kappa \lambda_{n}^{2}x) \Big].
\end{equation}
A very commonly used expression in the literature for the Gaussian model is obtained from the solution of the advection-diffusion equation (Eq. \eqref{eq 7} with $\alpha=1$) in an infinite medium. For this equation to satisfy the boundary conditions given by equations \eqref{eq 8} and \eqref{eq 9}, with $-\infty < x< \infty$ and $-\infty < z< \infty$, a 'mirror-image' source is considered. The solution for this so called operational Gaussian model (O-G model) is \cite{Csanady}
\begin{equation}\label{eq 16}
\overline{c^{y}}(x,z)=\frac{Q}{2\sqrt{\pi \kappa x }u}\Big[\exp\Big(-\frac{(z-H_{s})^{2}}{4 \kappa x}\Big)+\exp \Big(-\frac{(z+H_{s})^{2}}{4 \kappa x}\Big) \Big].
\end{equation}
In the next section we are going to compare the solution \eqref{eq 13} of our fractional model \eqref{eq 7} against experimental data and both the Gaussian \eqref{eq 15} and O-G \eqref{eq 16} solution models.

\section{Results and discussion}

In order to analyze the performance of the fractional model proposed in this work, in comparison to the Gaussian model derived from Fick's law, we consider only the simplest case of constant values for wind speed and eddy diffusivity. The average wind speed is obtained directly from the experimental data. To obtain a constant eddy diffusivity $K_z$, we follow the procedure introduced in \cite{Goulart2017}, and we consider a spatial average $K_z=\langle K\rangle$ of a eddy diffusivity that is a linear function of downwind distance expressed by  $K=\rho u x$, where $\rho$ is the turbulence parameter. The turbulence parameter $\rho$ is parameterized as the square of turbulent intensity using Taylor statistical theory of diffusion $\rho=(\frac{\sigma_{w}}{u})^{2}$ \cite{Arya}, where $\sigma_{w}$ is the standard deviation of the vertical wind speed component. In addition, to investigate the potential of application of fractional differential equations to model real data, we separated the experimental data from the Copenhagen experiment \cite{Gryning1984}, the Prairie Grass experiment \cite{Barad} and the Hanford experiment \cite{Nickola} into two groups: one with $\frac{h}{|L|}<10$ and another with $\frac{h}{|L|}>10$, where $L$ is the Monin-Obukhov length \cite{Obukhov} and $h$ is height of PBL. The parameter $\frac{h}{|L|}$ is obtained from the energy balance equation in a turbulent flow. This parameter can be used to evaluate some characteristics of the physical structure of the turbulent flow. For $\frac{h}{|L|}<10$ we have a predominance of mechanical energy input (wind shear) in the turbulent flow. For $\frac{h}{|L|}>10$ we have a predominance of energy input by thermal convection in the turbulent flow. The analysis of these two situations aims to show that the value of $\alpha$ (order of the fractional equation) that best describes the concentration distribution, when confronting the proposed fractional model with the experimental data, is related to the physical structure of the turbulent flow. Usually, the performance of dispersion models is evaluated from a well know set statistical indices described by Hanna \cite{Hanna} defined in the following way,
\begin{equation*}
\begin{split}
\text{NMSE}\; (\text{normalized mean square error})&= \frac{\overline{(c_{o}-c_{p})^{2}}}{\overline{c_{o}}\overline{c_{p}}},\\
\text{Cor}\; (\text{correlation coefficient})&= \frac{\overline{(c_{o}-\overline{c_{p}})(c_{p}-\overline{c_{p}})}}{\sigma_{o}\sigma_{p}},\\
\text{FB}\; (\text{fractional bias})&=\frac{\overline{c_{o}-\overline{c_{p}}}}{0.5(\overline{c_{o}}+\overline{c_{p}})},\\
\text{FS}\; (\text{fractional standard deviations}) &= \frac{\sigma_{o}-\sigma_{p}}{0.5(\sigma_{o}+\sigma_{p})},
\end{split}
\end{equation*}
where $c_{p}$ is the computed concentration, $c_{o}$ is the observed concentration, $\sigma_{p}$ is the computed standard deviation, $\sigma_{o}$ is the observed standard deviation, and the overbar indicates an averaged value. The statistical index FA2 represents the fraction of data for $0.5 \leq\frac{c_{p}}{c_{o}}\leq 2 $. The best results are indicated by values nearest to $0$ in NMSE, FS and FB, and nearest to $1$ in Cor and FA2.

\subsection{Results for $\frac{h}{|L|}<10$ }

\begin{figure}\caption{Scatter diagram of observed and predicted concentration for the Copenhagen experiment (instable).}\label{fig1}
\includegraphics[width=0.5\textwidth]{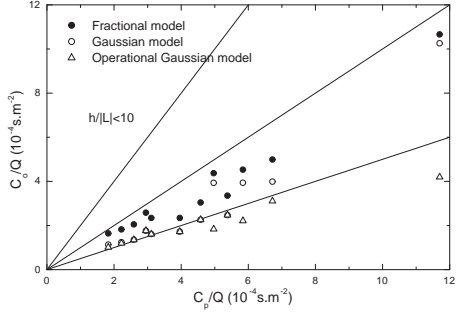}
\end{figure}

In order to estimate the better $\alpha$ value for each experiment, we analyzed the solution of our model from $\alpha=0.60$ to $\alpha=0.99$ by steps of $0.01$. An important result we found is that, regardless of the experiment, for $\frac{h}{|L|}<10$ the model (with constant velocity and constant eddy diffusivity) describes relatively well all experiments with $\alpha = 0.72$.
\begin{table}[!h]
\centering
\caption{\small Copenhagen Experiment for $\frac{h}{|L|}<10$ (instable)}\label{tab1}
\begin{tabular}{lccccccc}
\hline\hline
   Model              &      Cor      &     NMSE           &                    FS             &      FB                 &    FA2        \\
\hline\hline
  Eq.\eqref{eq 13}   &      0.97     &     0.05           &                    0.08           &     -0.24               &    1.00       \\
  Gauss               &      0.96     &     0.17           &                    0.06           &     -0.44               &    0.75       \\
 O-G                   &      0.97     &     0.83           &                    1.00           &     -0.77               &    0.41       \\
 Moreira (2005)       &      0.97     &     0.02           &                    0.05           &      0.01               &    1.00		\\
 Kumar (2012)        &      0.90     &     0.05           &                    0.34           &     -0.04               &    0.96       \\
  \hline\hline
\end{tabular}
\end{table}
In this case, for the Copenhagen experiment, Table \ref{tab1} shows that, when compared to the Gaussian and Operational Gaussian (O-G) models using an average wind speed and eddy diffusivity, our model generates good results for the concentration distribution of contaminants in the PBL generated by a turbulent flow where the source of thermal convection and mechanical input is quite relevant. It also presents good results when compared to some Eulerian dispersion models found in the literature, which employ integer-order derivatives in the advection-diffusion equation (traditional integer-order models are derived from Fick's law). In the case of the Moreira  model \cite{Moreira2005b}, the stationary advection-diffusion equation employs a wind speed that is a function of height $z$, and a eddy diffusivity that is a function of height $z$ and horizontal distance $x$. The differential equation is solved by the GILTT method \cite{Wortmann}, extended to the case where the eddy diffusivity is a function of $z$ and $x$. In the case of the Kumar model \cite{Kumar} the GILTT method is used to solve the advection-diffusion equation, but in this case, the wind speed and eddy diffusivity are functions only of the height $z$.
Table \ref{tab1} shows that the fractional model proposed in this work, where an average wind speed and eddy diffusivity (therefore constant) was used, has a similar performance to the model of \cite{Moreira2005b}. We also see that the fractional model proposed in this work performs better than Kumar model \cite{Kumar}, which is also a traditional model (derived from the integer-order), but considers a wind speed and a eddy diffusivity dependent on the height $z$. We can also observe that a wind speed and eddy diffusivity that more correctly describes the turbulent flow, used in the models of Moreira \cite{Moreira2005b} and Kumar \cite{Kumar}, tends to compensate the deficiency of the mathematical structure of the classical advection-diffusion equation to describe the concentration distribution. In our work, a constant wind speed and eddy diffusivity were used precisely to show the ability of the fractional advection-diffusion equation to more accurately describe the concentration distribution of contaminants.
In addition to the statistical indices, Figure \ref{fig1} shows the scatter diagram of observed concentration and predicted concentration. Lines indicate a factor of two.

\begin{figure}\caption{Scatter diagram of observed and predicted concentration for the Prairie Grass experiment (stable). (a) $x=200m$  and (b) $x=800m$}\label{fig2}
\includegraphics[width=0.49\textwidth]{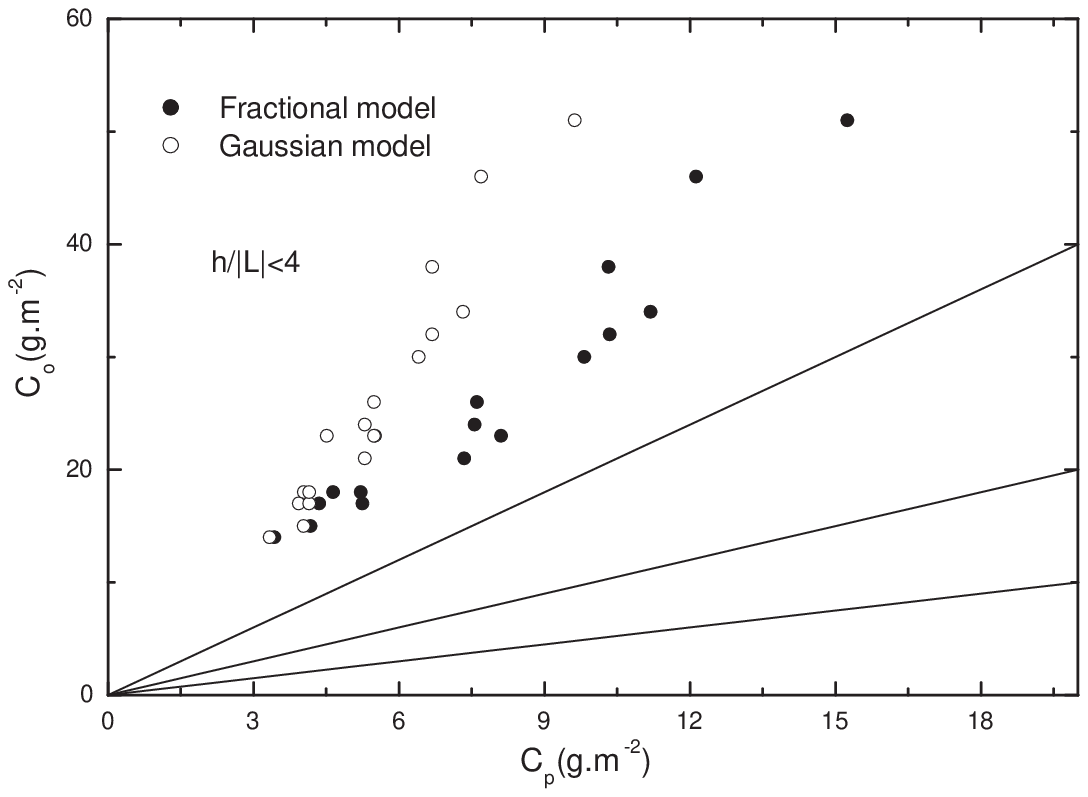}
\includegraphics[width=0.49\textwidth]{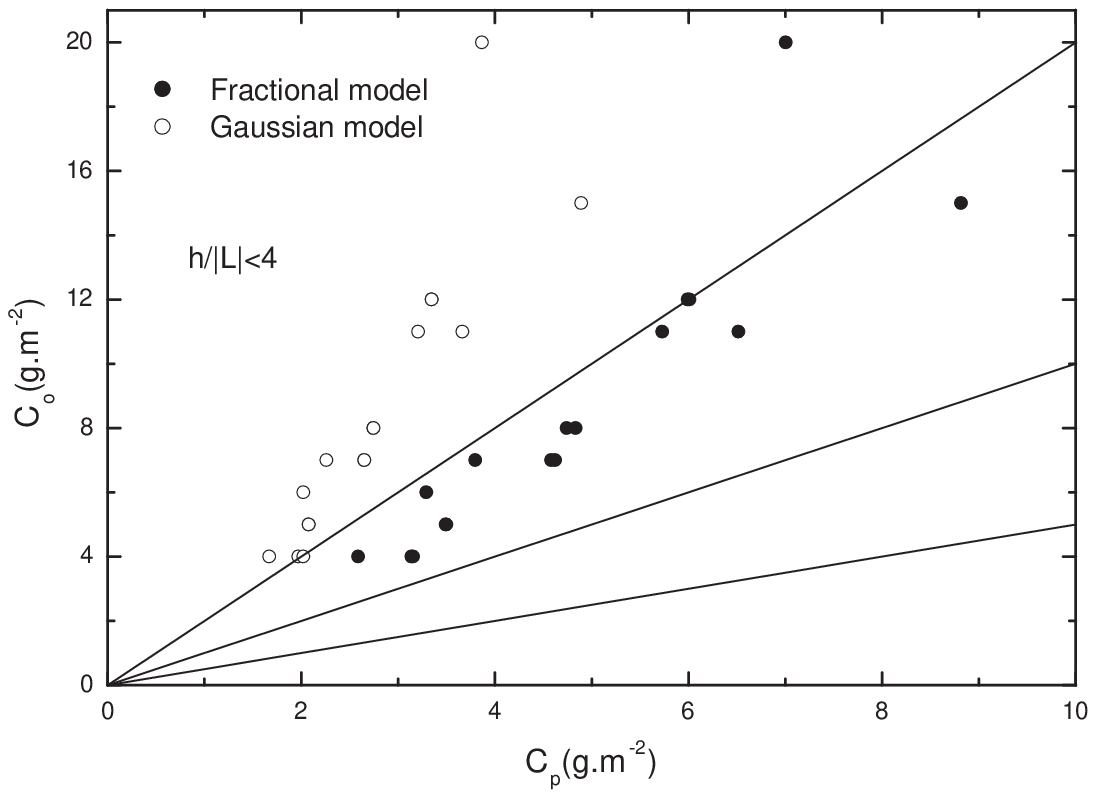}
\end{figure}

In Table \ref{tab2} we show the results obtained for the stable Prairie Grass experiment with $\frac{h}{|L|}<4$. In this case, the stability regime can be considered closer to neutral. We can observe that the fractional model with $\alpha=0.72$ performs better than both Gaussian and O-G models for all distances. In addition, we see that for greater distances the difference between the fractional model and the Gaussian model is more evident. The Figure \ref{fig2} shows the scatter diagram of observed concentration and predicted concentration for 200m and 800m distances.
We can observe that the performance of the fractional model is superior to the performance of the Gaussian model at both distances. We attribute the better performance of the fractional model in relation to the Gaussian model to the fact that the dispersion of contaminants in a turbulent flow does not obey a Gaussian probability distribution but rather an anomalous diffusion distribution, induced by the turbulence, with power law mean square displacement \cite{Metzler2000}.

\begin{table}[!h]
\centering
\caption{\small Prairie Grass Experiment for $\frac{h}{|L|}<4$ (stable)} \label{tab2}
\begin{tabular}{lccccccc}
\hline\hline
   Distance (m)           &      Model                 &     Cor            &           NMSE             &      FS                &    FB        &    Fa2       \\
\hline\hline
  \multirow{3}{*}{50}                      &      Eq.\eqref{eq 13}     &     0.96           &           4.39             &     1.22               &    -1.47     &    0.00      \\
                        &      Gauss                 &     0.96           &           4.91             &     1.53               &    -1.47     &    0.00     \\
                       &      O-G                    &     0.96           &           4.91             &     1.53               &    -1.47     &    0.00     \\ \hline
 \multirow{3}{*}{200}                  &      Eq.\eqref{eq 13}     &     0.96           &           1.68             &     1.06               &    -1.08     &    0.00      \\
                       &      Gauss                 &     0.96           &           0.95             &     1.08               &    -0.83     &    0.06     \\
                       &      O-G                    &     0.96           &           0.95             &     1.08               &    -0.83     &    0.06     \\ \hline
 \multirow{3}{*}{800}                     &      Eq.\eqref{eq 13}     &     0.90           &           0.47             &     0.89               &    -0.56     &    0.88      \\
                      &      Gauss                 &     0.84           &           1.72             &     1.35               &    -1.02     &    0.06     \\
                      &      O-G                    &     0.84           &           1.72             &     1.35               &    -1.02     &    0.06     \\ \hline
  \multirow{3}{*}{$\geq 200 $}             &      Eq.\eqref{eq 13}     &     0.91           &           1.35             &     1.20               &    -0.89     &    0.41      \\
             &      Gauss                 &     0.97           &           2.82             &     1.46               &    -1.20     &    0.06     \\
             &      O-G                    &     0.97           &           2.82             &     1.46               &    -1.20     &    0.06     \\ \hline
  \multirow{3}{*}{$\geq 400 $}             &      Eq.\eqref{eq 13}     &     0.93           &           0.73             &     1.03               &    -0.68     &    0.41      \\
             &      Gauss                 &     0.95           &           2.06             &     1.40               &    -1.08     &    0.08     \\
             &      O-G                    &     0.95           &           2.06             &     1.40               &    -1.08     &    0.08     \\ \hline
  \multirow{3}{*}{All distances}           &      Eq.\eqref{eq 13}     &     0.85           &           3.72             &     1.47               &    -1.27     &    0.31      \\
           &      Gauss                 &     0.98           &           4.56             &     1.54               &    -1.39     &    0.02     \\
           &      O-G                   &     0.98           &           4.56             &     1.54               &    -1.39     &    0.02     \\

   \hline\hline
\end{tabular}

\end{table}

\begin{table}[!h]
\centering
\caption{\small Hanford Experiment for $\frac{h}{|L|}\leq 3$ (stable)}\label{tab3}
\begin{tabular}{lccccccc}
\hline \hline
   Distance (m)            &      Model                 &     Cor            &           NMSE             &      FS                &    FB        &    Fa2      \\
\hline \hline
  \multirow{3}{*}{100}                      &      Eq.\eqref{eq 13}      &     0.65           &           10.32            &     1.80               &    -1.66     &    0.00     \\
                      &      Gauss                 &     0.63           &           13.21            &     1.84               &    -1.71     &    0.00     \\
                      &      O-G                    &     0.62           &           15.33            &     1.88               &    -1.74     &    0.00     \\ \hline
  \multirow{3}{*}{1600}                     &      Eq.\eqref{eq 13}      &     0.97           &           0.36             &     0.72               &    -0.53     &    1.00     \\
                     &      Gauss                 &     0.97           &           0.42             &     0.69               &    -0.58     &    0.66     \\
                     &      O-G                    &     0.88           &           6.21             &     1.78               &    -1.49     &    0.00     \\ \hline
  \multirow{3}{*}{3200}                     &      Eq.\eqref{eq 13}      &     0.95           &           0.10             &     0.50               &    -0.19     &    1.00     \\
                     &      Gauss                 &     0.95           &           0.12             &     0.49               &    -0.22     &    1.00     \\
                     &      O-G                    &     0.84           &           6.54             &     1.81               &    -1.48     &    0.00     \\ \hline
  \multirow{3}{*}{$\leq 800 $}              &      Eq.\eqref{eq 13}      &     0.60           &           8.27             &     1.77               &    -1.51     &    0.05     \\
              &      Gauss                 &     0.48           &           10.32            &     1.79               &    -1.58     &    0.00     \\
              &      O-G                    &     0.82           &           14.24            &     1.83               &    -1.71     &    0.00     \\ \hline
  \multirow{3}{*}{$\geq 1600 $}             &      Eq.\eqref{eq 13}      &     0.92           &           0.26             &     0.67               &    -0.38     &    1.00     \\
             &      Gauss                 &     0.90           &           0.29             &     0.65               &    -0.42     &    0.83     \\
             &      O-G                    &     0.83           &           6.31             &     1.75               &    -1.49     &    0.00     \\ \hline
  \multirow{3}{*}{All distances}            &      Eq.\eqref{eq 13}      &     0.55           &           7.57             &     1.74               &    -1.35     &    0.43     \\
            &      Gauss                 &     0.39           &           9.31             &     1.76               &    -1.42     &    0.33     \\
            &      O-G                    &     0.87           &           14.01            &     1.80               &    -1.68     &    0.00     \\

   \hline \hline
\end{tabular}
\end{table}

\begin{figure}\caption{Scatter diagram of observed and predicted concentration for the Hanford experiment. (a) $x=800m$ and (b) $x=3200m$}\label{fig3}
\includegraphics[width=0.49\textwidth]{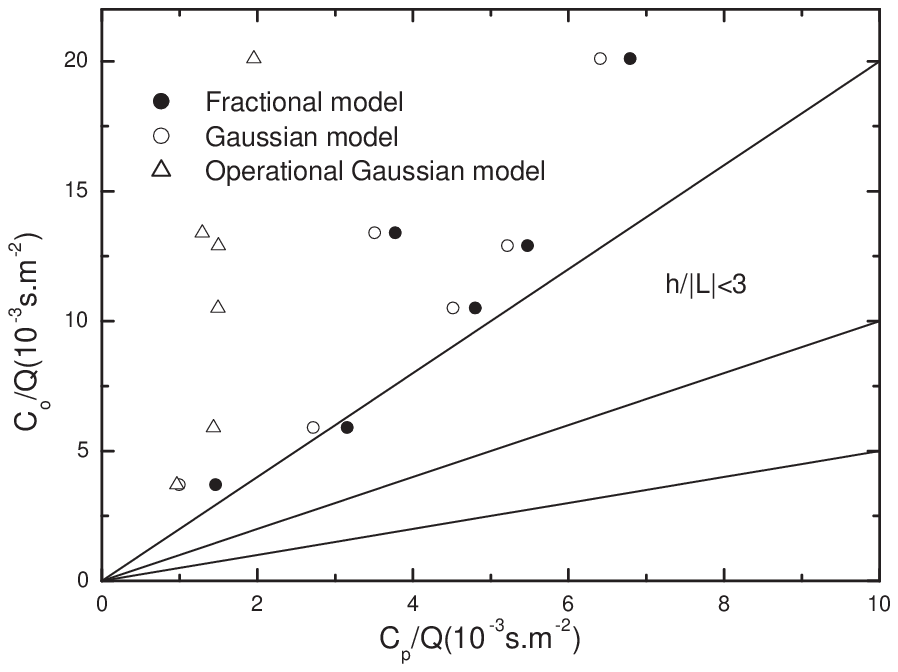}
\includegraphics[width=0.49\textwidth]{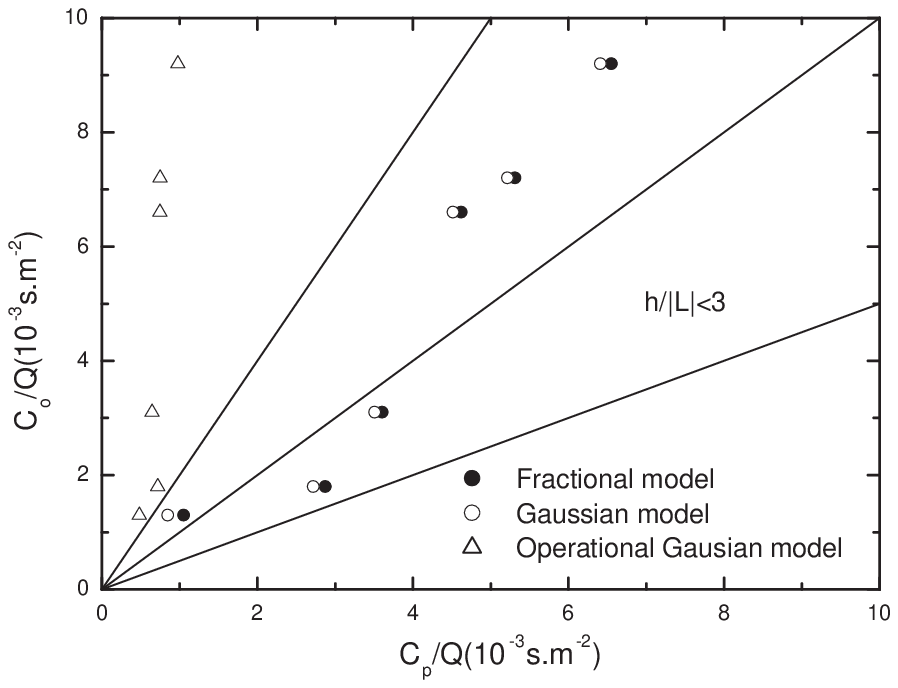}
\end{figure}

Table \ref{tab3} shows the results of the fractional, Gaussian and O-G models for the Hanford experiment with $\frac{h}{|L|}<10$. Again, the fractional model has a better performance than both Gaussian and O-G models. However, the difference between the fractional and the Gaussian model is smaller than in the cases of the Copenhagen and Prairie Grass experiments. The experiment of Copenhagen took place in an unstable regime where the source of thermal convection and mechanical input is quite relevant. In this case, the flow is strongly turbulent.  The Prairie Grass experiment occurred in a stable regime, but with a high wind speed ($\approx 5.3 ms^{-1}$). In this case, the mechanical source of the turbulence is strong. Furthermore, in the Prairie Grass experiment, small distances from the source are considered. It is reasonable to suppose that a diffusion model with constant wind speed and constant eddy diffusivity, that is, a model that does not completely describe the physical structure of the turbulent flow, performs better in a turbulent flow where the properties vary little with vertical height and for large distances from the source, where the dispersion of the contaminant can be described by the statistical properties of the turbulent flow. In the Hanford experiment, we have a stable regime, with a very low wind speed ($\approx 1.5 m s^{-1}$). Therefore, the flow is slightly turbulent and the characteristics of anomalous diffusion, present in a strongly turbulent flow, are not very evident. In this case, it is expected that the Gaussian probability distribution can be employed to model the problem. Figure \ref{fig3} shows the scatter diagram of observed concentration and predicted concentration for the Hanford experiment. We can see that the fractional and Gaussian models perform very similarly. However, the O-G model, given by equation \eqref{eq 16}, fails completely in this case. This is related to the approximation made so that the solution satisfies the boundary conditions \eqref{eq 8} and \eqref{eq 9}. The values of the distances in the $z$ and $x$ directions involved in the experiments of Copenhagen and Prairie Grass are similar in almost all experiments. In this case, the approximation of boundary at infinity gives a good result. In particular, for the stable Prairie Grass where the $z$ and $x$ lengths are $\approx 400m$ and $800m$, respectively, the difference between Gaussian and O-G is negligible. In the Hanford experiment, the values of the distances in the $z$ and $x$ directions are very different ($ \approx 200 m$ and $3200m$, respectively), making the boundary at infinity approach totally inadequate.

\subsection{Results for $\frac{h}{|L|}>10$ }

\begin{table}[!h]
\centering
\caption{\small Copenhagen Experiment for $\frac{h}{|L|}>10$ (instable)}\label{tab4}
\begin{tabular}{lccccccc}
\hline \hline
   Model              &      Cor      &     NMSE           &                    FS             &      FB                 &    FA2        \\
\hline \hline
  Eq.\eqref{eq 13}    &      0.65     &     0.20           &                    0.97           &     -0.14               &    0.90       \\
  Gauss               &      0.62     &     0.34           &                    1.02           &     -0.33               &    0.81       \\
 O-G                   &      0.71     &     0.47           &                    1.09           &     -0.47               &    0.72       \\
 \hline \hline
\end{tabular}
\end{table}

\begin{table}[!h]
\centering
\caption{\small Prairie Grass Experiment for $\frac{h}{|L|}>10$ (instable)}\label{tab5}
\begin{tabular}{lccccccc}
\hline \hline
  Distance (m)            &      Model                 &     Cor            &           NMSE             &      FS                &    FB       &    Fa2      \\
\hline \hline
  \multirow{3}{*}{50}                      &      Eq.\eqref{eq 13}      &     0.80           &           0.37             &     -1.04              &    0.07     &    0.75     \\
                      &      Gauss                 &     0.89           &           0.40             &     -0.76              &    0.58     &    0.75     \\
                      &      O-G                    &     0.89           &           0.40             &     -0.76              &    0.58     &    0.75     \\ \hline
  \multirow{3}{*}{100}                     &      Eq.\eqref{eq 13}      &     0.48           &           1.26             &    -1.59               &    0.66     &    0.65     \\
                     &      Gauss                 &     0.41           &           1.24             &    -1.38               &    0.91     &    0.20     \\
                     &      O-G                    &     0.41           &           1.24             &    -1.38               &    0.91     &    0.20     \\
  \hline \hline
\end{tabular}
\end{table}

For $\frac{h}{|L|} > 10$, Tables \ref{tab4} and \ref{tab5} show that the proposed fractional model  with $\alpha = 0.80$ performs better when compared to the Gaussian model in most statistical indices. As in the previous case, we analyzed the solutions of our model from $\alpha=0.60$ to $\alpha=0.99$ by steps of $0.01$, and we found that the model (with constant wind speed and constant eddy diffusivity) describes relatively well all experiments with $\alpha = 0.80$ when $\frac{h}{|L|} > 10$. We also observed that the performance of all models with $\frac{h}{|L|} > 10$ is worst than the performance when $\frac{h}{|L|} < 10$ (Tables \ref{tab1}, \ref{tab2} and \ref{tab3}). This may be related to the difference in the structure of the turbulent flow in the PBL in both cases. When $\frac{h}{|L|} > 10$ a free-convection-like state emerges \cite{Wyngaard}.  In this case, we have a strong turbulence generated by thermal convection and a great variation in the structure of the flow in the vertical direction. This results in a large variation with the height of the intensity of the vertical eddy diffusivity. As in the experiments of Prairie Grass (unstable), and cases 1, 3, 7 and 8 of the Copenhagen experiment. On the other hand, when $\frac{h}{|L|}$ is very small ($\frac{h}{|L|} < 10$) there is a turbulence where the mechanical source (wind shear) is relevant. In this case, we have a greater spatial homogenization in the flow. Consequently, we have a low variation with the height of the intensity of the vertical eddy diffusivity \cite{Plein}, as in the Prairie Grass (stable), and the Hanford experiments, and also in cases 2, 4, 5, 6 and 9 of the Copenhagen experiment.
Models that employ a constant wind speed and eddy diffusivity will have greater difficulty in correctly describing the distribution concentration when there is a greater spatial asymmetry, as in the case of the PBL flow if $\frac{h}{|L|} > 10$. This deficiency will be naturally compensated when non-constant wind speed and eddy diffusivity are used to more accurately describe the physical structure of the flow. However, in the present work we consider only constant wind speed and eddy diffusivity since, at this moment, we are interested only in demonstrating that the structure of the fractional order differential equation is more appropriate to calculate the distribution of dispersed contaminants in a turbulent flow than an integer-order differential equation. We are confronting a Gaussian model with fractional order derivatives against the Gaussian model with integer-order derivatives. Of course, a fractional or an integer-order model which considers a non-constant wind speed and eddy diffusivity, that more adequately describes the structure of the turbulent flow, will generate more realistic concentration distribution values. However, in general, it is not simple to obtain an analytical solution for both equations (fractional and integer-order) to easily compare its performances.

Finally, the most important result we found by considering both $\frac{h}{|L|} < 10$ and $\frac{h}{|L|} > 10$ is that there should be a relation between the order of the fractional derivative with the physical structure of the turbulent flow since, regardless of the experiment, when we have a predominance of mechanical energy input ($\frac{h}{|L|} < 10$) in the turbulent flow all experiments where best described with $\alpha = 0.72$, and when we have a predominance of energy input by thermal convection ($\frac{h}{|L|}>10$) the experimental data is better described by $\alpha = 0.80$.


\section{Conclusions}

In the present work, we propose a new parameterization for the flow of concentration using fractional derivatives. The fractional order differential equation in the longitudinal and vertical directions is used to obtain the concentration distribution of contaminants in the Planetary Boundary Layer. The use of fractional derivatives in modeling diffusion of contaminants is justified by the presence of anomalous diffusion generated from turbulence. In the last decades, thousands of works were carried out with the objective of explaining anomalous diffusion. However, in this subject only a few works dealt with the analysis of the validity of models based on classical differential equations, and/or the use of unusual differential operators, to describe systems displaying non-differential behavior and/or anomalous dynamics. In this context, the use of fractional differential operators emerged as a valuable mathematical tool to model the time evolution of anomalous diffusion \cite{Metzler2000, Metzler2004, West}. However, the use of fractional calculus to study steady-state regimes of dispersion of contaminants in the atmosphere is actually underexplored. In this respect, we recently demonstrated in \cite{Goulart2017} that the steady-state regime for a spatial concentration distribution of a non-reactive contaminants, in an anomalous diffusion process displaying a power law mean square displacement, is given naturally by a differential equation containing fractional derivatives in the advective term ($x$ direction) of the advection-diffusion equation. In the present work we propose a new parameterization for the flux of concentration using fractional derivatives (Eq. \eqref{eq 5}). In this case the resulting fractional  advection-diffusion equation containing fractional derivatives in the advective and diffusive terms ($x$ and $z$ directions), with this the model consider anomalous diffusion in both directions, differently of our previous fractional model \cite{Goulart2017} where the anomalous diffusion was considered only in the longitudinal direction. We solve the model and we compare the solution with both real experiments and traditional integer order derivative models. We show that our fractional model, even with constant wind speed and a constant eddy diffusivity, gives very good result in fitting the experimental data. The results obtained show that the structure of the fractional order differential equation is more appropriate to calculate the distribution of dispersed contaminants in a turbulent flow than an integer-order differential equation. Furthermore, a very important result we found it is that there should be a relation between the order $\alpha$ of the fractional derivative with the physical structure of the turbulent flow since, regardless the experiment, when we have a predominance of mechanical energy input in the turbulent flow all experiments where best described with $\alpha = 0.72$, and when we have a predominance of energy input by thermal convection the experimental data is better described by $\alpha = 0.80$. We are currently working towards a mathematical relationship between some parameters that describe the physical structure of the flow with the value of $\alpha$. Finally, the result obtained from the fractional derivative model motivates further investigations of applications of fractional differential equations to model diffusion of contaminants. There are many directions of investigation left to explore. Interesting examples include a theoretical investigation of a direct relationship between the fractal structure of turbulent flow and the power law mean square displacement for the anomalous diffusion (namely, a direct relation between the fractal structure and the order $\alpha$ of the fractional derivative), the use of local fractional/fractal derivatives in the model instead of the non-local Caputo derivatives, a detailed analyzes of the spatial memory effect introduced by a non-local derivative in the model, etc. These and other examples are left to future works.


\section*{Acknowledgments}

This work was supported in part by CNPq and CAPES, Brazilian funding agencies.

\appendix

\section{Solution of the fractional model}\label{appendix}

In this appendix we obtain the solution \eqref{eq 13} of the fractional differential equation \eqref{eq 7} subjected to the boundary conditions \eqref{eq 8} and \eqref{eq 9}.
The fractional differential equation \eqref{eq 7} can be analytically solved by using the separation of variables technique. We begin by assuming that the solution can be written as follows,
\begin{equation}\label{eq A1}
\overline{c^{y}}(x,z)=X(x)Z(z).
\end{equation}
By substituting \eqref{eq A1} into \eqref{eq 7}, we get two ordinary fractional differential equations in the variables $X$ and $Z$ as follows,
\begin{equation}\label{eq A2}
{_0^C D^{\alpha}_x} X+\kappa \lambda^{2} X=0
\end{equation}
and
\begin{equation}\label{eq A3}
\frac{d }{d z}[{_0^C D^{\alpha}_z} Z]+\lambda^{2} Z=0.
\end{equation}
The solution for the system of equations \eqref{eq A2} and \eqref{eq A3} can be obtained by Frobenius method. The Eq. \eqref{eq A2} has solution $X(x) = E_{\alpha}(-\kappa\lambda^2 x^\alpha)$, \cite{Goulart2017}. In order to solve \eqref{eq A3} by Frobenius method, we consider the following power series
\begin{equation}\label{eq A4}
\begin{array}{lll}
 Z(z^\alpha)=z^{p}\displaystyle \sum_{j=0}^{\infty}b_{j}\frac{z^{j (\alpha+1)}}{\Gamma( (\alpha+1)j+\beta)},
\end{array}
\end{equation}
where $0<\alpha<1$, and $p$, $\beta$ are constants. The Caputo fractional derivative of \eqref{eq A4} is then given by
\begin{equation}\label{eq A5}
\begin{array}{lll}
 {_0^C D^{\alpha}_z} Z(z^{\alpha})&  = &\frac{b_{0}}{\Gamma(\beta)}  {_0^C D^{\alpha}_z} z^{p}  + \displaystyle\sum_{j=1}^{\infty}\frac{b_{j}}{\Gamma( (\alpha+1)j+\beta)} \frac{\Gamma((\alpha+1)j +p+1)}{\Gamma( (\alpha+1)j +p-\alpha+1)}z^{(j-1) \alpha + j + p}.
\end{array}
\end{equation}
By considering $\beta = p + 1$ we get
\begin{equation}\label{eq A6}
\begin{array}{lll}
 {_0^C D^{\alpha}_z} Z(z^{\alpha}) = \frac{b_{0}}{\Gamma(p+1)}D_{z}^{\alpha} z^{p} + \displaystyle\sum_{j=1}^{\infty}\frac{b_{j}}{\Gamma( (\alpha+1)j +p-\alpha+1)}z^{(j-1) \alpha + j + p},
\end{array}
\end{equation}
where (Goulart et al., 2017),
\begin{equation}\label{eq A7}
{_0^C D^{\alpha}_z} z^{p} = \left\{\begin{array}{cl}
& 0, \;\;\;\;\;\;\;\;\;\;\;\;\;\;\;\;\;\;\;\;\;\;\;\;\; p=0 \\
& \frac{\Gamma (p+1)}{\Gamma (p+1-\alpha)} z^{p-\alpha}, \;\;\;\;\;\;\;\;0 < p \leq 1
\end{array} \right. .
\end{equation}
Then we have
\begin{equation}\label{eq A8}
\begin{array}{lll}
\frac{d }{d z} {_0^C D^{\alpha}_z} Z(z^{\alpha}) =  \frac{b_{0}}{\Gamma(p+1)}\frac{d }{d z} {_0^C D^{\alpha}_z} z^{p} + \displaystyle\sum_{j=0}^{\infty}\frac{b_{j+1}}{\Gamma((\alpha+1)j +1+p)}z^{j \alpha + j + p},
\end{array}
\end{equation}
where, we use the identity $\Gamma(x+1) = x\Gamma(x)$.

From \eqref{eq A7} we have two cases when $0<p\leq 1$ and $p=0$, respectively.

\subsection*{Case $0<p\leq 1$.}

For the case $0<p\leq 1$ we have from \eqref{eq A8}
\begin{equation}\label{eq A9}
\begin{array}{lll}
\frac{d }{d z} {_0^C D^{\alpha}_z} Z(z^{\alpha}) =   \frac{b_{0}(p-\alpha)}{\Gamma (p+1-\alpha)} z^{p-1-\alpha} + \displaystyle\sum_{j=0}^{\infty}\frac{b_{j+1}}{\Gamma( (\alpha+1)j +1+p)}z^{(\alpha +1) j + p}.
\end{array}
\end{equation}
By substituting equation \eqref{eq A9} into \eqref{eq A3} we obtain
\begin{equation}\label{eq A10}
\begin{array}{lll}
 \frac{b_{0}(p-\alpha)}{\Gamma (p+1-\alpha)} z^{p-1-\alpha} + \displaystyle\sum_{j=0}^{\infty}\frac{1}{\Gamma( (\alpha+1)j +1+p)}\Big(b_{j+1} + \lambda^2 b_{j}\Big)z^{(\alpha +1) j + p}=0,
\end{array}
\end{equation}
that yields $p=\alpha$ and $b_{j+1} +\lambda^2 b_{j} = 0$. Then the solution of \eqref{eq A3} when we have $0<p\leq 1$ is given by
\begin{equation}\label{eq A11}
Z_1(z) = b_0z^\alpha\displaystyle\sum_{j=0}^{\infty}\frac{(-\lambda^2)^j z^{(\alpha + 1)j}}{\Gamma((\alpha+1)j +\alpha + 1)} = b_0z^\alpha E_{\alpha +1,\alpha+1}(-\lambda^2 z^{\alpha + 1} ).
\end{equation}
But this solution does not satisfy the boundary condition $ {_0^C D^{\alpha}_z} Z_1(0) = 0$ because
\begin{equation}\label{eq A12}
\begin{array}{lll}
 {_0^C D^{\alpha}_z} Z_1(z) = \frac{b_{0}}{\Gamma(1)}   + b_0z^\alpha \displaystyle\sum_{j=1}^{\infty}\frac{(-\lambda^2)^j}{\Gamma(j (\alpha+1) +1)}z^{ (\alpha +1)j -\alpha },
\end{array}
\end{equation}
then, if $b_0\neq 0$,
\begin{equation}\label{eq A13}
\begin{array}{lll}
 {_0^C D^{\alpha}_z} Z_1(0) = \frac{b_{0}}{\Gamma(1)} \neq 0.
\end{array}
\end{equation}
Therefore $Z_1(z)$ does not gives the solution for our problem.

\subsection*{Case $p = 0$.}

For the case $p = 0$ we have from \eqref{eq A8},
\begin{equation}\label{eq A14}
\begin{array}{lll}
\frac{d }{d z} {_0^C D^{\alpha}_z} Z(z^{\alpha}) =   \displaystyle\sum_{j=0}^{\infty}\frac{b_{j+1}}{\Gamma( (\alpha+1)j +1)}z^{j \alpha + j}.
\end{array}
\end{equation}
By substituting equation \eqref{eq A14} into \eqref{eq A3} we obtain
\begin{equation}\label{eq A15}
 \displaystyle\sum_{j=0}^{\infty}\frac{1}{\Gamma( (\alpha+1)j +1)}\Big(b_{j+1} +\lambda^2 b_{j}\Big)z^{j \alpha + j} = 0,
\end{equation}
that yields $b_{j+1} +\lambda^2 b_{j} = 0$. Then, the solution of \eqref{eq A3} in this case is given by
\begin{equation}\label{eq A16}
Z_2(z) = b_0\displaystyle\sum_{j=0}^{\infty}\frac{(-\lambda^2)^j z^{(\alpha + 1)j}}{\Gamma((\alpha+1)j +1)} = E_{\alpha +1}(-\lambda^2 z^{\alpha + 1} ),
\end{equation}
that gives the solution for our problem. From \eqref{eq A16} we have
\begin{equation}\label{eq A17}
\begin{array}{lll}
 {_0^C D^{\alpha}_z} Z_2(z) =b_0  \displaystyle\sum_{j=1}^{\infty}\frac{(-\lambda^2)^j}{\Gamma( (\alpha+1) j-\alpha+1)}z^{ (\alpha +1)j - \alpha }.
\end{array}
\end{equation}
By applying the boundary conditions \eqref{eq A9} we get
\begin{equation}\label{eq A18}
\begin{array}{lll}
 {_0^C D^{\alpha}_z} Z_2(0) =0,
\end{array}
\end{equation}
and
\begin{equation}\label{eq A19}
\begin{array}{lll}
 {_0^C D^{\alpha}_z} Z_2(h) =b_0  \displaystyle\sum_{j=1}^{\infty}\frac{(-\lambda^2)^j}{\Gamma( (\alpha+1) j-\alpha+1)}h^{(\alpha+1)j - \alpha } = 0.
\end{array}
\end{equation}
Consequently, whe should have $\lambda = \lambda_n$, where $\lambda_n$ is a solution for the equation
\begin{equation}\label{eq A20}
\begin{array}{lll}
(\lambda^2 h) \displaystyle\sum_{j=0}^{\infty}\frac{(-\lambda^2 h^{\alpha+1})^j}{\Gamma( (\alpha+1) j+2)} = 0.
\end{array}
\end{equation}

Finally, from the superposition principle we obtain \eqref{eq 13}:

\begin{equation}\nonumber
\overline{c^{y}}(x,z)=\sum_{n=0}^{\infty}a_{n}E_{\alpha}(-\kappa \lambda_{n}^{2}x^{\alpha})E_{\alpha+1}(-\lambda_{n}^{2}z^{\alpha+1}).
\end{equation}

In order to fix the constants $a_n$ we use the boundary condition $ c^{y}(0,z)=\frac{Q}{u}\delta(z-H_{s})$ in \eqref{eq 13}. We consider
\begin{equation}\label{eq A21}
a_0+\sum_{n=1}^{m}a_{n}E_{\alpha+1}(-\lambda_{n}^{2}z^{\alpha+1})=\frac{Q}{u}\delta(z-H_{s})
\end{equation}
as a good approximation because, for $z\neq 0$, $E_{\alpha+1}(-\lambda_{n}^{2}z^{\alpha+1})$ goes to zero when $n$ goes to infinite. Now, by making
\begin{equation}\label{eq A22}
\int^h_0( a_0+\sum_{n=1}^{m}a_{n}E_{\alpha+1}(-\lambda_{n}^{2}z^{\alpha+1}))E_{\alpha+1}(-\lambda_{p}^{2}z^{\alpha+1})dz=\frac{Q}{u}\int^h_0 \delta(z-H_{s})E_{\alpha+1}(-\lambda_{p}^{2}z^{\alpha+1})dz,
\end{equation}
where $p\in \{0,1,2,...,m\}$, we fix $a_n$ from the numerical solution of the following system:
\begin{equation}\label{eq A23}
\int^h_0( a_0+\sum_{n=1}^{m}a_{n}E_{\alpha+1}(-\lambda_{n}^{2}z^{\alpha+1}))E_{\alpha+1}(-\lambda_{p}^{2}z^{\alpha+1})dz=\frac{Q}{u}E_{\alpha+1}(-\lambda_{p}^{2}H_s^{\alpha+1}).
\end{equation}



\begin{thebibliography}{99}

\bibitem{Metzler2000}Metzler, R. and Klafter, J., 2000. The random walk's guide to anomalous diffusion: a fractional dynamics approach. Physics Reports 339, 1 -- 77.

\bibitem{Metzler2004}Metzler, R. and Klafter, J., 2004. The restaurant at the end of the random walk: recent developments in the description of anomalous transport by fractional dynamics. J. Phys. A: Math. Gen. 37, R161.

\bibitem{Richardson}Richardson, L.F., 1926. Atmospheric diffusion shown on a distance-neighbour graph. Proc. R. Soc. Lond. A. 110, 709 -- 737.

\bibitem{West}West, B.J., 2014. Fractional calculus view of complexity: A tutorial. Rev. Mod. Phys. 86, 1169.

\bibitem{West2}West, B.J., 2015. Fractional Calculus View of Complexity: Tomorrow's Science. CRC Press, Taylor \& Francis Group. Boca Raton.

\bibitem{Magin}Magin, R.L., Abdullah, O., Beleanu, D., Zhou, X.J., 2008. Anomalous diffusion expressed through fractional order differential operators in gthe Bloch--Torrey equation. Journal of Magnetic Renonance 190, 255 -- 2701.

\bibitem{Gomez-Aguilar}Gomez-Aguilar, J.F., Miranda-Hernandez, M., Lopez-Lopez, M. G., Alvarado-Martinez, V.M. , Baleanu, D., 2016. Modeling and simulation of the fractional space-time diffusion equation. Commun. Nonlinear Sci. Numer. Simul. 30, 115--127.

\bibitem{Mandelbrot}Mandelbrot, B.B., 1982. The Fractal Geometry of Nature. W. H. Freeman and Company. New York.

\bibitem{Gryning1987}Gryning, S.E., Holtslag, A.M.M., Irwin, J.S., Sivertsen, B., 1987. Applied dispersion modelling based on meteorological scaling parameters. Atmos. Environ. 21, 79 -- 89.

\bibitem{Hanna and Paine}Hanna, S.R. and Paine, R.J., 1989. Hybrid plume dispersion model (hpdm)  development and evaluation. J. Appl. Meteorol. 28, 206 -- 224.

\bibitem{Moreira2005a}Moreira, D.M., Rizza, U., Vilhena, M.T., Goulart, A., 2005. Semi-analytical model for pollution dispersion in the planetary boundary layer. Atmos. Environ. 39, 2689 -- 2697.

\bibitem{Wilson}Wilson, J.D., Sawford, B.L., 1996. Review of Lagrangian stochastic models for trajectories in the turbulent atmosphere. Boundary-Layer Meteorol. 78, 191 -- 210.

\bibitem{Taylor}Taylor, G.I., 1921. Diffusion by continuous movements. Proc. London Math. Soc. 20, 196 -- 212.

\bibitem{Batchelor} Batchelor, G.K., 1949. Diffusion in a field of homogeneous turbulence, Eulerian analysis. Australian Journal of Science Research 2, 437 -- 450.

\bibitem{Degrazia}Degrazia, G.A., Anfossi, D., Carvalho, J. C., Mangia, C, Tirabassi, T., 2000. Turbulence parameterization for pbl dispersion models in all stability conditions.   Atmos. Environ. 34, 3575 -- 3583.

\bibitem{Goulart2004}Goulart, A., Moreira, D.M., Carvalho, J.C., Tirabassi, T., 2004. Derivation of eddy diffusivities from an unsteady turbulence spectrum. Atmos. Environ. 38, 6121 -- 6124.

\bibitem{Goulart2017}Goulart, A.G.O., Lazo, M.J., Suarez, J. M. S., Moreira, D.M., 2017. Fractional derivative models for atmospheric dispersion of pollutants. Physica A.  477,  9--19.

\bibitem{Gryning1984}Gryning, S.E., Lyck, E., 1984. Atmospheric dispersion from elevated source in an urban area: comparison between tracer experimental and model calculations. Journal of Climate and Applied Meteorology. 23, 651 -- 654.

\bibitem{Barad} Barad, M.L., 1958. Project Prairie Grass: A Field Program in Diffusion. Geophysical Research Paper No. 59, Vols. I and II, AFCRL-TR-58-235 (ASTIA Document No. AF-15572), Air Force Cambridge Research Laboratories, Bedford, England.

\bibitem{Nickola}Nickola, P.W.,  1977. The Hanford 67-Series: a volume of atmospheric diffusion measurements. PNL-2433, Battelle,Pacific Northwest Laboratory, Richland, USA.

\bibitem{Pasquill}Pasquill, F., Smith, F.B., 1983. Atmospheric Diffusion. Halsted Press, USA.

\bibitem{Tarasov} Tarasov, V. E., 2010. Fractional Dynamics: Applications of Fractional Calculus to Dynamics of Particles, Fields and Media. Springer-Verlag Berlin Heidelberg.

\bibitem{Csanady}Csanady, G.T., 1973. Turbulent Diffusion in the Environment. D. Reidel Publishing, Holland.

\bibitem{Seinfeld}Seinfeld, J.H., 1986. Atmospheric Chemistry and Physics of Air Pollution. Wiley-Interscience Publishing, USA.

\bibitem{Diethelm}Diethelm, K., 2010.  The Analysis of Fractional Differential Equations: An Application-Oriented Exposition Using Differential Operators of Caputo Type.  Springer-Verlag, Berlin Heidelberg.


\bibitem{Arya} Arya, S.P., 1995. Modeling and parameterization of near-source diffusion in weak winds. J. Appl. Meteorol. 34, 1112--1122.

\bibitem{Obukhov}Obukhov, A.M., Turbulence in an atmosphere with a non-uniform temperature. Boundary-Layer Meteorology. 2, 7 -- 29.

\bibitem{Hanna}Hanna, S.R., 1989. Confidence limit for air quality models as estimated by bootstrap and jacknife resampling methods. Atmos. Environ. 23, 1385--1395.

\bibitem{Moreira2005b}Moreira, D.M., Vilhena, M.T., Tirabassi, T., Buske, D., Cotta, R., 2005. Near-source atmospheric pollutant dispersion using the new GILTT method. Atmospheric Environment. 39 6289 -- 6294.

\bibitem{Wortmann}Wortmann, S., Vilhena, M.T., Moreira, D.M., Buske, D., 2005. A new analytical approach to simulate the pollutant dispersion in the PBL. Atmos. Environ. 39, 2171 -- 2178.

\bibitem{Kumar}Kumar, P. and Sharan, M., 2012. Parametrization of the eddy diffusivity in a dispersion model over homogeneous terrain in the atmospheric boundary layer. Atmospheric Research 106, 30 -- 43.

\bibitem{Wyngaard}Wyngaard, J.C., 2010. Turbulence in the Atmosphere. Cambridge University Press, Cambridge.

\bibitem{Plein}Pleim, J.E., Chang, J.S., 1992. A non-local closure model for vertical mixing in the convective boundary layer. Atmos. Environ. 26A, 965 -- 981.

\bibitem{Hentschel}Hentschel, H.G.E., Procaccia, I., 1982. Intermittency Exponent in Fractally Homogeneous Turbulence. Physical Review Letters. 49, 1158--1161.

\bibitem{Procaccia}Procaccia, I., 1984. Fractal Structures in Turbulence. Journal of Statistical Physics. 36 649--663.

\bibitem{Frisch}Frisch, U., Sulen, P., 1978. A simple dynamical model of intermittent fully developed turbulence. J. Fluid Mech. 87 719--738.

\end{thebibliography}
\end{document}